\documentclass[conference]{IEEEtran}
\IEEEoverridecommandlockouts
\usepackage{cite}
\usepackage{amsmath,amssymb,amsfonts}
\usepackage{algorithmic}
\usepackage{graphicx}
\usepackage{textcomp}
\usepackage{xcolor}
\usepackage{multirow}
\usepackage{booktabs}
\usepackage[numbers]{natbib}
\usepackage{bm}
\usepackage{amsmath}
\renewcommand{\baselinestretch}{1.00}
\def\BibTeX{{\rm B\kern-.05em{\sc i\kern-.025em b}\kern-.08em
    T\kern-.1667em\lower.7ex\hbox{E}\kern-.125emX}}
\begin{document}

\title{FPAENet: Pneumonia Detection Network Based on Feature Pyramid Attention Enhancement\\
}

\author{\IEEEauthorblockN{Xudong Zhang$^{1,2}$, Bo Wang$^{1,2}$, Di Yuan$^{1,2}$, Zhenghua Xu$^{1,2,\dag}$, Guizhi Xu$^{1,2}$}
	\IEEEauthorblockA{
		$^1$State Key Laboratory of Reliability and Intelligence of Electrical Equipment,\\Hebei University of Technology, China\\
		$^2$Key Laboratory of Electromagnetic Field and Electrical Apparatus Reliability of Hebei Province,\\Hebei University of Technology, China\\
		$^{\dag}$Corresponding author, email: zhenghua.xu@hebut.edu.cn}
}
\maketitle

\begin{abstract}
Automatic pneumonia Detection based on deep learning has increasing clinical value. Although the existing Feature Pyramid Network (FPN) and its variants have already achieved some great successes, their detection accuracies for pneumonia lesions in medical images are still unsatisfactory. In this paper, we propose a pneumonia detection network based on feature pyramid attention enhancement, which integrates attended high-level semantic features with low-level information. We add another information extracting path equipped with feature enhancement modules, which are conducted with an attention mechanism. Experimental results show that our proposed method can achieve much better performances, as a higher value of $4.02\%$ and $3.19\%$, than the baselines in detecting pneumonia lesions.

\end{abstract}


\section{Introduction}

Chest radiograph(CXR) is an important screening technology for patients with pulmonary disease, which is widely applied in pneumonia examination and tracking development~\cite{pinto2013scoring}~\cite{rajpurkar2017chexnet}. The computer-aided diagnosis based on deep learning for object detection aims to automatically detect the interested objects (e.g., lesions). Particularly, pulmonary disease detection based on deep learning is one of the most important tasks~\cite{zech2018variable, yan20183d, dai2016r}. FPN~\cite{lin2017feature} augments a standard convolutional network with a top-down pathway and lateral connections so the network efﬁciently constructs a rich, multi-scale feature pyramid from a single resolution input image. RetinaNet~\cite{lin2017focal} and EfficientDet~\cite{tan2020efficientdet} are two object detection networks with good performance at present, which have high detection speed and high detection accuracy. In~\cite{tan2020efficientdet}, Mingxing.T proposed the BiFPN, which increases bottom-up information flowing than FPN. And its classification and regression networks are the same as RetinaNet.
 
\begin{figure}
	\centering
	\includegraphics[scale=0.4]{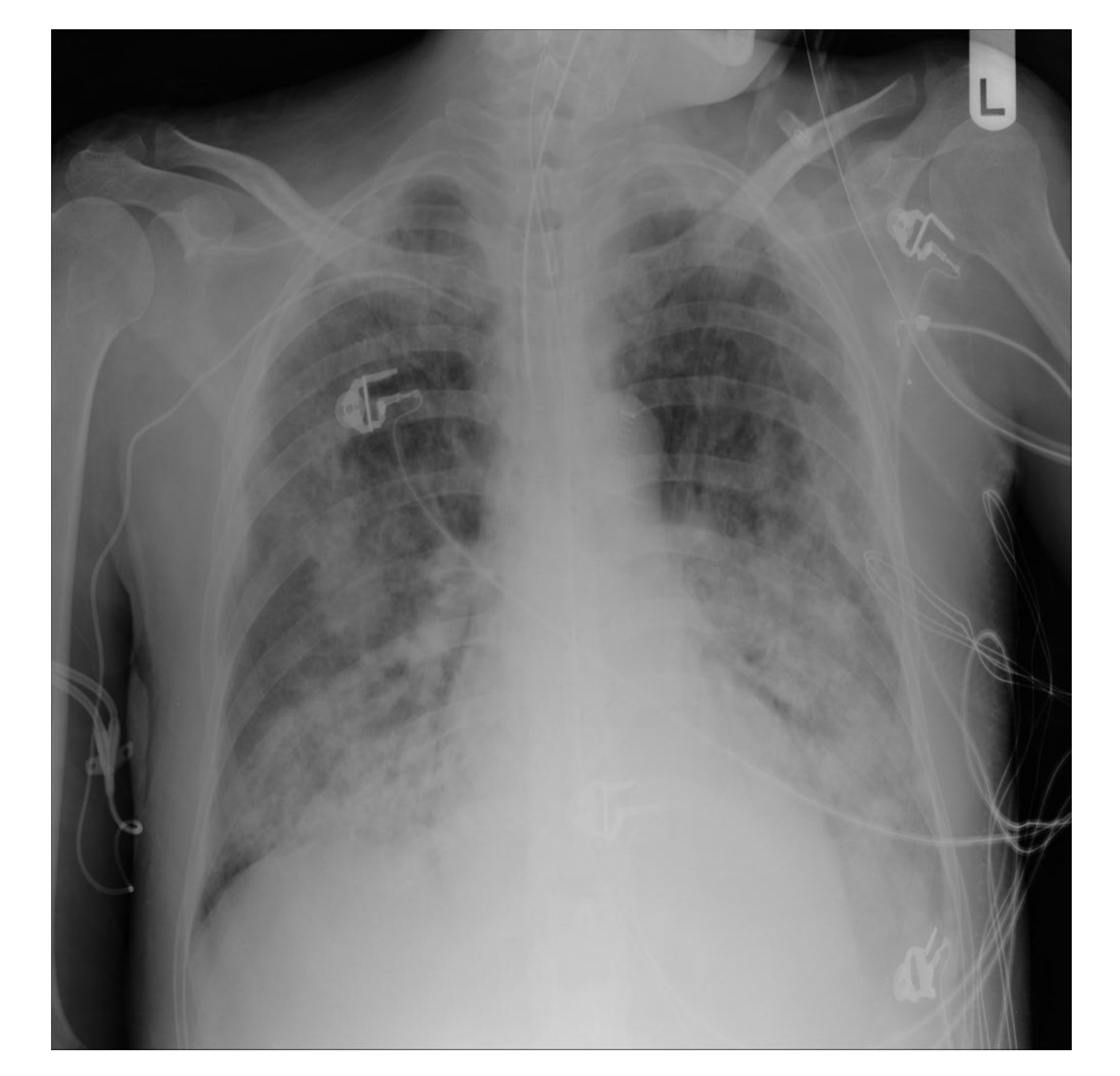}
	\caption{An example of pneumonia CXR images. And the lungs have begun to fibrosis.}
	\label{FIG:1}
\end{figure} 
 
Although FPN and its variants have already achieved some great successes, their detection accuracies for pneumonia lesion in medical images are still unsatisfactory. Specifically, in the context of medical images, the pneumonia lesions often relatively look like ground glass, as shown in Fig.~\ref{FIG:1}, which means that the fibrosis caused by pneumonia lesions are intermixed with background images~\cite{self2013high}, so that it is more difficult to detect the lesions in CXR images compared with other images. 
Moreover, the detection effect is limited, due to the difference in X-ray devices,quality of images,larger number of relevant diseases~\cite{ticinesi2016lung}.
Therefore, in the feature maps of every levels, the features of these unclear objects may be captured not enough, which thus results in inaccurate detection for pneumonia lesions. This inaccurate detection performance may lead to some severe consequences in clinical practice.

The contributions of this paper are briefly summarized as follows: 
\begin{itemize}
	\item We point out the challenge of  pneumonia lesions intermixed with background image, and propose a novel FPENet model to resolve this problem by fusing two top-down channels and feature enhancement.
	\item We also propose an attention in feature enhancement to increase the weight of classification and location feature, which further improves the deep model’s detection capability.
	\item Extensive experiments are conducted on a public pneumonia dataset, the results show that our proposed method can achieve much better performances than the baselines in detecting pneumonia lesions. In addition, ablation studies show that both feature pyramid enhancement and attention modules are essential for FPAENet to achieve such superior detection performances.
\end{itemize}

The rest of the paper is organized as follows. In Section~\ref{Related Work}, we briefly review previous studies on object detection. In Section~\ref{Method}, we introduce our proposed  FPAENet respectively. In Section~\ref{Experiment}, our proposed FPAENet method is evaluated and compared with the current detection network. In addition, the components and parameters of our network are analyzed in detail.
In Section~\ref{Discussion}, We discuss our proposed FPAENet and the baseline model.
The paper is finally concluded in Section~\ref{Conclusion}. 

\section{Related Work}
\label{Related Work}
In this section, we briefly review some previous works, including the two-stage,one-stage and some other methods for medical images specifically.

With the development of computer vision, the algorithm of object detection is developing fastly and used widely. The object detection algorithm can be divided into one-stage, two-stage and method for medical images specifically.  

$\bm {Two-Stage.}$
In~\cite{girshick2014rich}, Ross Girshick proposed R-CNN that is the first two-stage algorithm. It started with the extraction of a set of object proposals by the selective search (SS). Then each proposal is rescaled to a fixed size image and fed into the convolutional neural networks(CNN) that trained on ImageNet to extract features. Finally, linear SVM~\cite{he2016deep} classifier was used to predict the presence of an object within each region and to recognize object categories. The Spatial Pyramid Pooling(SPP) layer of  Spatial Pyramid Pooling Networks (SPPNet) in~\cite{he2015spatial} enabled CNN to generate a fixed-length proposal, and SPPNet avoided repeatedly computing the feature maps with CNN. In~\cite{girshick2015fast}, R.Girshick proposed the Fast R-CNN that enabled us to simultaneously train a detector and a bounding box regressor under the same network configurations. Region Proposal Networks (RPN) is proposed in~\cite{ren2015faster} to produce higher quality boxes, especially as a single network. The concept of the anchor was put forward in RPN, which was the boxes of different scales and proportions for each pixel in the last layer of the feature map firstly, then corrects the position of the box through bounding regression, and determines whether the box is foreground or background. Finally, the filtered boxes are classified and regressed through RoI pooling. Compared with SS, RPN is not only faster but also of higher quality. However, due to the existence of the RPN network, the speed of the network is relatively slower than that of the one-stage. RFCN reduces computation and improves model speed by increasing Shared information and introducing location-sensitive score maps before ROI pooling. 
Although the two-stage model obtained higher accuracy than the one-stage, the speed of the two-stage was slower than the one-stage.

$\bm{Method\quad for\quad Medical\quad Images.}$
3DCE precisely adopted the structure of RFCN. Unlike RFCN, 3DCE required feature extraction from multiple adjacent CT slices. 3DCE's approach was to divide every three adjacent images into a group, then extracted the features of each group, and finally concatenated them together.
In~\cite{shao2019attentive,tao2019improving,yan2019mulan,li2019mvp,yan2018deeplesion}, several variants have been proposed, including adding attention and feature fusion. In addition, 3DCE and its variants only use the characteristics of the intermediate core slice as the input of the RPN to obtain the proposal, and then classifies and returns after the position-sensitive region of interest (PSROI). The limitation of the method can't be ignored. It required the fusion of slice so that it can't be applied in 2D images.

$\bm {One-Stage.}$
Compared with the higher accuracy of Two-stage, one-stage had a faster speed because it needed RPN to produce the candidate boxes. The series of $YOLO$~\cite{redmon2016you,redmon2017yolo9000,redmon2018yolov3, bochkovskiy2020yolov4} and SSD~\cite{liu2016ssd} had a good performance in speed as the main and representative algorithm of one-stage.
Due to the idea of anchor, each pixel of the feature map had several anchors corresponding to the original image, and most of the proposals were negative examples and relatively easy to classify. A large number of negative examples account for a large proportion of the loss function, which made the loss shift to the negative examples during backpropagation and limited the detection ability of the model. Tsung-Yi.L et.al proposed Focal Loss, which reduces the loss of easy-to-discriminate negative examples, so that the loss can better optimize the parameters during backpropagation. Meanwhile, RetinaNet was proposed, which still uses the anchor method, adopts the FPN architecture, and used two parallel FCN for classification and regression behind each layer of feature maps. 
Among them, Focal Loss was used to reduce the imbalance between positive and negative cases. For all one-stage models, the quality of the feature information obtained before classification and regression was crucial to classification and regression. BiFPN was proposed that can obtain high-quality feature information by increasing the information fusion of different layers. 

In this paper,  we mainly follow the one-stage detector design, and we show it is possible to achieve higher accuracy with improved network architectures.

\begin{figure*}
	\centering
	\includegraphics[scale=0.7]{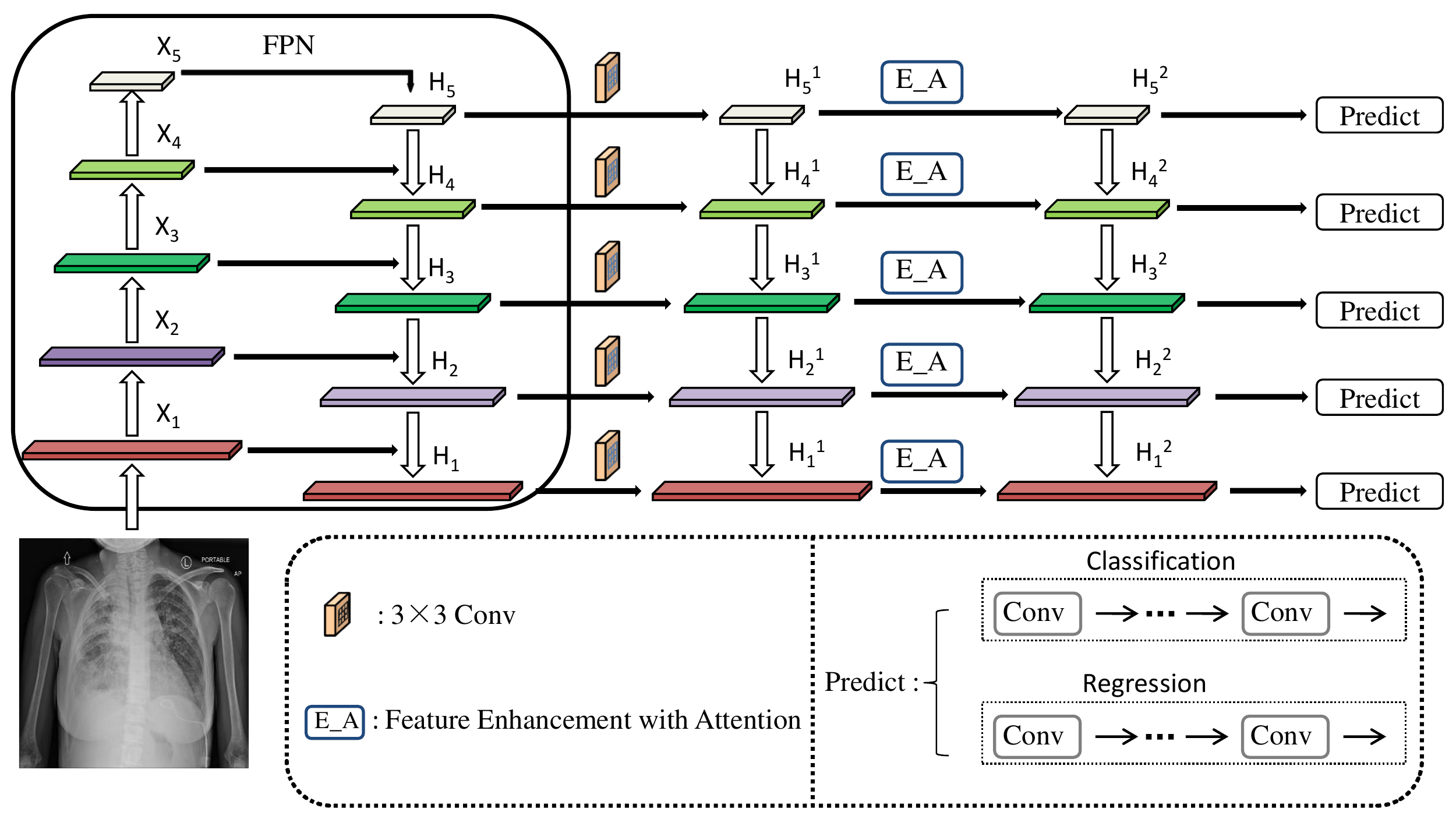}
	\caption{The framework of our proposed FPAENet method. ResNet-50 as the backbone to extract features. Two top-down channels are added in the FPN, and feature enhancement with attention is placed on the horizontal connection to enhance the effective information. Next, two parallel Fully Convolutional Networks to classify whether the candidate area is a lesion and locate the lesion. }
	\label{FIG:3}
\end{figure*}

\section{Method}

\label{Method}
In this part, we introduce in detail our proposed FPAENet method, including the architecture of our network (Section~\ref{Network Architecture}), and another two important modules: Feature Enhancement (Section~\ref{Feature Enhancement}) and Attention (Section~\ref{Attention}).

\subsection{Network Architecture}
\label{Network Architecture}
The framework of our proposed FPAENet method is shown in Fig.~\ref{FIG:3}. In order to solve the problem of abundant background in the lesion area, we prppose the FPAENet to deal with this particularity of penumonia CXR images. 

We use the $ResNet-50$~\cite{he2016deep} as the backbone network to extract features, and finally get the feature map of the last five layers, denoted as   $X_i$ ($i \in \{ 1, 2 \cdots 5 \}$), whose size is $4\times4$, $8\times8$, $16\times16$, $32\times32$ and $64\times64$ respectively, and the number of channels is 256. According to the architecture of FPN, whose formula is shown as Equation~\ref{E1}.

\begin{equation}
\label{E2}
H_i^{1} = \varphi(H_{i}) + upsample(H_{i+1}^{1})
\end{equation}

\subsection{Feature Enhancement}
\label{Feature Enhancement}

The application of the convolution kernel of different sizes can not only extract features in depth but also realize the fusion of features in a wider range.
Through the addition of features, feature information is strengthened, and attention will be used to regulate the proportion of significant information. 
These operations enable the detector to extract more accurate and information-enhanced features from CXR images of lungs that its target areas are riching in background. 

\begin{figure}
	\centering
	\includegraphics[scale=0.47]{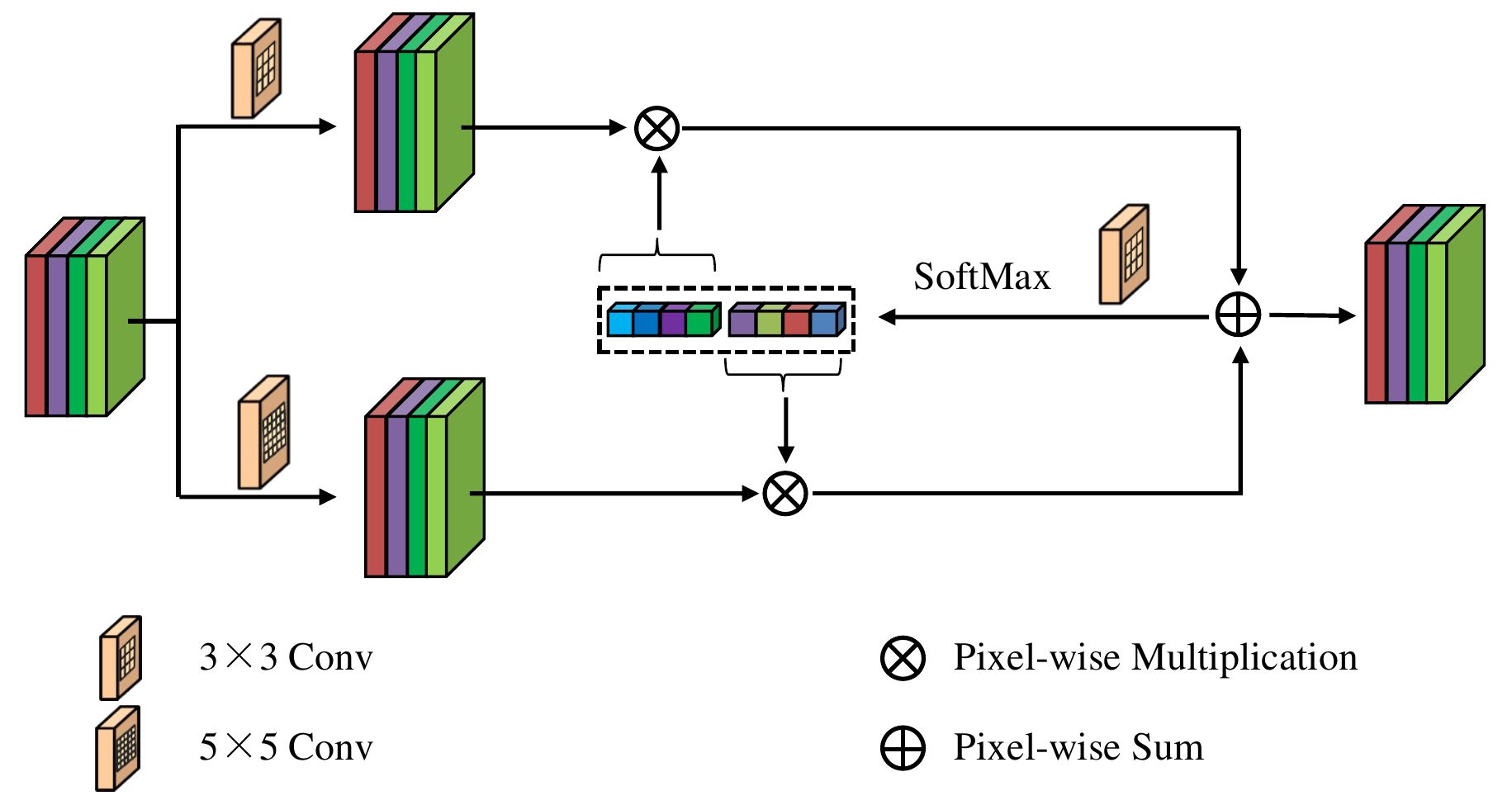}
	\caption{The framework of feature enhancement with attention in our proposed FPAENet method.}
	\label{FIG:4}
\end{figure}

\subsection{Attention}
\label{Attention}
The attention module aims to selectively aggregate features from input images by attending to the relevant context information among different layers.

\begin{equation}
\label{E8}
w=SoftMax(A) 
\end{equation}

With this attention module, the features from diﬀerent layers are attentively aggregated with a learnable cross-slice attention vector to amplify the relevant contextual features and suppress irrelevant ones.

\section{Experiment}
\label{Experiment}
In this section, we first introduce the dataset and compare our proposed FPAENet method with the current mainstream detection network. Then, we validate the effectiveness of the important components of our method, include the two top-down channels, the enhancement of feature, and attention. After that, we further evaluate the influence of the network parameters(e.g., the thresholds of IOU).

\subsection{Experimental Settings}

The proposed method was validated in the dataset of the lung of RSNA, which contains a total of 6012 CXR images and 1019 of them were used as the testing dataset to test the effectiveness of our method. 
In the training process, we used $ResNet-50$ pre-trained on ImageNet as the feature extraction backbone to extract features. The epoch of the training out model is 10. The batch size is 2. Setting the learning rate to be 0.00001 with Adam \cite{kingma2014adam} as the optimizer. The threshold of IoU is 0.5, and we trained and tested in NVIDIA GeForce GTX 2080Ti GPUs.

And we used the mAP as the measure to measure the effectiveness of our method. 
\begin{equation}
mAP = \dfrac{\sum_{i=1}^{K} AP_i}{K}
\end{equation}
K is the number of classes. In this dataset, $K = 1$. And $P$ stands for accuracy, whose calculation method is shown in Equation~\ref{9}.
\begin{equation}
\label{9}
P = \dfrac{TP}{TP + FP}
\end{equation}

\subsection{Main Results}

The proposed FPAENet method will be compared with RetinaNet and EfficientDet that both of them had achieved good results on the dataset of COCO. We trained and tested the proposed FPAENet method, RetinaNet, and EfficientDet that are reproduced on the same dataset. The result of mAP obtained by the competing methods(i.e, RetinaNet, EfficientDet) and our FPAENet method are presented in Table~\ref{tab1}.

\begin{table}[!h]
	\centering
	\small
	\caption{\upshape The result of our proposed FPAENet method and baseline models (i.e, RetinaNet, EfficientDet) tested in the same dataset.}%
	\begin{tabular}{c|c|c}
		\toprule
		\textbf{Method} & \textbf{Backbone} & \textbf{mAP} \\
		\midrule
		\multirow{ 2}{*}{RetinaNet}
		& {ResNet-50} & {$45.49\%$} \\
		& {ResNet-101} & {$46.93\%$} \\
		\hline
		\multirow{ 2}{*}{EfficientDet}
		& {ResNet-50} &  { $46.38\%$ } \\
		& {ResNet-101} & {$41.33\%$} \\
		\hline
		\multirow{ 2}{*}{\textbf{FPAENet}}
		& {ResNet-50} & \bm{ $49.51\%$} \\
		& {ResNet-101} & {$45.17\%$} \\
		\bottomrule
	\end{tabular}\label{tab1}
\end{table}

Finally, when the $ResNet-50$ as the backbone, the detection effect of RetinaNet, EfficientDet, and our proposed FPAENet method is thus improved. However, when the $ResNet-101$ as the backbone, RetinaNet has the highest mAP, the mAP of our proposed FPAENet method is higher than EfficientDet. The reason for this may be because our data volume is small, and the deeper $ResNet-101$ requires more data to train.

In order to verify the effectiveness of the various modules of the proposed model, we did the following experiments, and the experimental results are shown in Table~\ref{tab2}.

\begin{table}[!h]
	\centering
	\caption{\upshape The results of mAP for singe mudule and overlay of modules. }%
	\begin{small}
		\begin{tabular}{c|c|c|c}
			\toprule
			\textbf{New Channels} & \textbf{Enhancement} & \textbf{Attention} & \textbf{mAP}  \\
			\midrule
			{$\surd$} & {} & {} & { $47.77\%$ } \\
			\hline
			{$\surd$} & {$\surd$} & {} & {$48.64\% $} \\
			\hline
			{$\surd$} & {$\surd$} & {$\surd$} & { $49.51\%$ } \\
			\bottomrule
		\end{tabular}\label{tab2}
	\end{small}
\end{table}

\section{Discusion}
\label{Discussion}
In this section, we first summarize our proposed FPAENet method. After that, we also print out the main differences between our proposed FPAENet and the baseline model(i.e.,  RetinaNet, EfficientDet). Finally, we briefly state the advantages of our proposed method.
\subsection{Summary on Our Proposed Method}
The method FPAENet proposed by us is to add two top-down channels that connected by horizontal convolution and enhance the feature with attention on the basis of FPN. Two parallel full-coil networks are followed by the feature map of each layer for classification and regression respectively. 
\subsection{Comparison with Baseline}
Compared with Retinanet using FPN directly, our improvement based on FPN is more suitable for the particularity of pneumonia CXR images. And different from the BiFPN adopted by EfficientDetect to strengthen the fusion of different layer semantics, our method enhanced the feature information while integrating the higher-level semantics, which improved the proportion of classification and location feature.
\subsection{Advantages}
Our proposed FPAENet method has a high detection accuracy on the images that the target area rich in background information. In the case of the same speed and memory consumption, Our proposed FPAENet method has a better detection capability than EfficientDet.

\section{Conclusion and Future Work}
\label{Conclusion}
In this work, in order to solve the peculiarities of a rich background in the target area of pneumonia CXR images, FPAENet was proposed to locate the lesion accurately, by increasing two top-down channels connected in horizontal and enhancing feature with attention. On the public dataset, the effectiveness of our proposed method on detecting lesions had been extensively evaluated. Compared with the current mainstream detection models (i.e, RetinaNet, EfficientDet), our proposed method improved the mAP by $4.02\%$ and $3.19\%$, respectively.

\section*{Acknowledgment}
This work was supported by the National Natural Science Foundation of China under the grant 61906063, by the Natural Science Foundation of Tianjin City, China, under the grant 19JCQNJC00400, by the ``100 Talents Plan'' of Hebei Province under the grant E2019050017, and by the Yuanguang Scholar Fund of Hebei University of Technology, China.

\bibliographystyle{IEEEtranS}
\bibliography{refs}

\end{document}